\documentclass{article}

\usepackage{epsfig}
\usepackage{here}

\newcommand{\markfont}{\small\sf}

\begin{document}
\newcommand{\fuss}[1]{{#1}}
\setcounter{page}{939}
\title{
\vspace*{-20mm}\mbox{\mbox{} \hspace*{24mm}\normalsize Published in: \underline{Lecture Notes
 in Computer Science {\bf 2415}, 939-944 (2002).}}\\[4mm] 
Asymptotic Level Density of the Elastic Net
 Self-Organizing Feature Map}
\author{Jens Christian Claussen and Heinz Georg Schuster
\\
\small
Institut f\"ur Theoretische Physik und Astrophysik
\\\small
Leibnizstr. 15, 24098 Christian-Albrechts-Universit\"at zu Kiel, Germany
\\
\small
\small \texttt{http://www.theo-physik.uni-kiel.de/\~{ }claussen/} \\[-3mm] } \date{}

\maketitle
\begin{abstract}
Whileas the Kohonen 
Self Organizing Map shows an asymptotic level density
following a power law with a magnification exponent 2/3, 
it would be desired to have an exponent 1 in order to
provide optimal mapping in the sense of
information theory.
In this paper, we study analytically and numerically
the magnification behaviour
of the Elastic Net algorithm 
as a model for self-organizing feature maps.
In contrast to the Kohonen map the Elastic Net
shows no power law, but for onedimensional maps
nevertheless the density follows an universal magnification
law, i.e.{} depends on the local stimulus density only
and is independent on position and decouples from
the stimulus density at other positions.
\end{abstract}
\noindent
Self Organizing Feature Maps map an input space, such as the
retina or skin receptor fields, into a neural layer by
feedforward structures with lateral inhibition.
Biological maps show as defining properties
topology preservation, error tolerance,
plasticity (the ability of adaptation to changes in input space),
and self-organized formation by a local process, since the
global structure cannot be coded genetically.
The self-organizing feature map algorithm
proposed by Kohonen \cite{kohonen82} has become a successful model
for topology preserving primary sensory processing
in the cortex \cite{obermayer92},
and an useful tool in technical applications \cite{ritter92}.

The Kohonen algorithm for Self Organizing Feature Maps
is defined as follows:
Every stimulus ${\bf v}$ of an euclidian input space $V$
is mapped to the neuron with the position ${\bf s}$
in the neural layer $R$
with the highest neural activity, given by the condition
\begin{eqnarray}
\mbox{$|{\bf w}_{{\bf s}} -{\bf v}|$} =
\mbox{$\min_{{\bf r}\in R}
|{\bf w}_{{\bf r}} -{\bf v}|$} \label{eq:voronoi}
\end{eqnarray}
where $| . |$ denotes the euclidian distance in input space.
In the Kohonen model the learning rule for each synaptic
weight vector ${\bf w}_{{\bf r}}$ is given by
        \begin{eqnarray}
        {\bf w}_{{\bf r}}^{{\sf new}} =
         {\bf w}_{{\bf r}}^{{\sf old}}
         + \eta \cdot g_{{\bf r} {\bf s}}
         \cdot ({\bf v}-{\bf w}_{{\bf r}}^{{\sf old}})
        \end{eqnarray}
with $g_{{\bf r}{\bf s}}$ as a gaussian function of euclidian
distance $|{\bf r}-{\bf s}|$ in the neural layer.
The function $g_{{\bf r}{\bf s}}$ describes the topology in the
neural layer. The  parameter $\eta$ determines the speed of
learning and can be adjusted during the learning process.
Topology preservation is enforced by the common
update of all weight vectors whose neuron ${\bf r}$ is adjacent
to the center of excitation ${\bf s}$.
\clearpage
\fuss{
\noindent \small \sf 940 \hspace*{0.3cm} J.C. Claussen and H.G. Schuster \hfill\normalsize\rm
}
\section{The Elastic Net Feature Map}
The Elastic Net \cite{durbin87} was proposed 
for solving optimization problems like the famous Travelling
Salesman Problem. Here we apply this concept 
to feature maps.
The Elastic Net is defined as a gradient
descent in the energy landscape
\begin{eqnarray}
E=-\sigma^2 \sum_{\mu} \ln \sum_r
{e}^{-({\bf v}^{\mu}-{\bf w}_r)^2/2\sigma^2}
  + \frac{\tilde{\kappa}}{2} \sum_r |{\bf w}_{r+1}-{\bf w}_{r}|^2
  \label{eq:elnetenergie}
\end{eqnarray}
with the input vectors denoted by ${\bf v}^{\mu}$.
Here $r$ is the index of the neurons in an one-dimensional array 
(for the TSP: with periodic boundary conditions), 
and ${\bf w}_{r}$ is the synaptic weight vector of that neuron.
For $\sigma\rightarrow 0$
(\ref{eq:elnetenergie}) becomes
\begin{eqnarray}
\lim_{\sigma\rightarrow{0}} E
= \frac{1}{2} \sum_{\mu}  
({\bf v}^{\mu}-{\bf w}_{s({\bf v}^{\mu})})^2
  + \frac{\tilde{\kappa}}{2} \sum_r |{\bf w}_{r+1}-{\bf w}_{r}|^2.
\end{eqnarray}
Here $s({\bf v}^{\mu})$ denotes the neuron with the smallest
distance to the stimulus, the winning neuron, which is
assumed to be nondegenerate.
A gradient descent in the first term (which can be interpreted
as an entropy term \cite{simic90})
leads for sufficiently small $\sigma$ to the condensation
of (at least) one weight vector to each input vector,
if the input space is discrete.
The second term is the potential energy of an elastic string
between the weight vectors, and gradient descent in this term
leads to a minimization
of the (squared!) distances between the weight vectors.

Depending on parameter adjustment 
\cite{durbin89,simmen91}
a gradient descent in $E$ can provide near-optimal solutions
to the TSP within polynomial processing time 
\cite{hertzkroghpalmer},
similar as the Kohonen algorithm \cite{ritter92}.
We remark that in the Travelling Salesman application
(if the numbers of neurons and cities are chosen to be equal)
both the Elastic Net and the
Kohonen algorithm share the same zero \cite{ritter92}
and first \cite{gruel92} order terms and
are therefore related for the final state of convergence,
although their initial ordering process is different.

The update rule of the Elastic Net Algorithm is the
gradient descent in~(\ref{eq:elnetenergie}):
\begin{eqnarray}
\frac{1}{\eta} \delta {\bf w}_{{\bf r}}
= \sum_{\mu} ({\bf v}^{\mu}-{\bf w}_{{\bf r}})
\frac{{e}^{-({\bf v}^{\mu}-{\bf w}_{{\bf r}})^2/2\sigma^2}
    }{\int {{d}}{\bf r}^{'}
      {e}^{-({\bf v}^{\mu}-{\bf w}_{{\bf r}^{'}})^2/2\sigma^2}}
  + \tilde{\kappa}
  \mbox{\boldmath$\bigtriangleup$\unboldmath}
  {\bf w}_{{\bf r}},       \label{eq:elnetgrad}
\end{eqnarray}
where $\mbox{\boldmath$\bigtriangleup$\unboldmath}
w_r = w_{r-1} - 2 w_r + w_{r+1}$
denotes the discrete Laplacian.

If we apply this concept to feature maps, we have to replace
the sum over all input vectors by an integral over
$\int p({\bf v}){\rm d}{\bf v}$, i.e.{} a probability density.
If we interpret (\ref{eq:elnetgrad})
as a neural feature mapping algorithm,
it is a pattern parallel learning rule, or batch update rule,
where contributions of all patterns are summed up to one update term.
In the brain, hovever, patterns are presented serially in a stochastic
sequence.
Therefore we 
generalize this algorithm
to serial presentation:
\begin{eqnarray}
\frac{1}{\eta}
\delta {\bf w}_{{\bf r}} &=& ({\bf v}-{\bf w}_{{\bf r}})
\frac{{e}^{-({\bf v}-{\bf w}_{{\bf r}})^2/2\sigma^2}
    }{\int {{d}}{\bf r}^{'}
      {e}^{-({\bf v}-{\bf w}_{{\bf r}^{'}})^2/2\sigma^2}}
      + \kappa
\mbox{\boldmath$\bigtriangleup$\unboldmath}
     {\bf w}_{{\bf r}}.
\label{eq:elnetseriell}
\end{eqnarray}
\clearpage
\fuss{
\noindent \small \markfont   \hfill  Asymptotic Level Density of the Elastic Net \hspace*{0.3cm} 941 \normalsize\rm \\ \\
}
\noindent In Monte Carlo simulations of this model, one chooses input vectors
${\bf v}$ according to the probablility density function
$p({\bf v})$ and updates ${\bf w}_{{\bf r}}$ for every neuron
${\bf r}$ in the neural layer according
to~(\ref{eq:elnetseriell}).
The algorithm can be viewed as a stochastic approximation
algorithm that converges if the conditions
$
\sum_{t=0}^{\infty} \eta^2(t) < \infty
$
and
$
\sum_{t=0}^{\infty} \eta(t) = \infty
$ 
for the time development of parameter $\eta$
are fulfilled \cite{kohonen91}
The simultaneous adjustment of $\kappa$ and $\sigma$ has been
discussed in 
\cite{durbin89,simmen91}
for the special case
of the TSP optimization problem. For the TSP it appears necessary to 
adjust $\kappa/\sigma$ to a system-size-dependent value
to avoid 'spike defects' for small $\kappa/\sigma$ 
and 'frozen bead defects' for large  $\kappa/\sigma$ when
annealing $\sigma\to{}0$ \cite{simmen91}.
Both 'defects' are no defects in feature maps, the 'spike defects'
can only occur for delta-peaked stimuli (cities) together with 
a dimension-reduction.

The aim in feature maps is different.
Using the Kohonen algorithm, one tries to start with large-ranged
interaction in the neural layer to avoid global topological defects.
This is not directly possible for the Elastic Net, as its 
learning cooperation is restricted to next-neighbour.
Only the strength of the elastic spring $\kappa$ can be initialized with
a high value and decreased after global ordering.
The parameter $\sigma$ is to be interpreted as a resolution length
in feature space, e.~g. the distance between two receptors in skin
or retina. For selectivity of the winner-take-all mechanism, one 
would choose $\sigma$ smaller or alike the average or minimal distance
between adjacent weight vectors.

\section{Asymptotic Density and the Magnification Factor}
In this paper we consider the case of continuously
distributed input spaces with same dimensionality as the neural
layer, so there is no reduction of dimension.

The magnification factor is defined as the density of
neurons ${\bf r}$ (i.~e. the density of
synaptic weight vectors ${\bf w}_{{\bf r}}$)
per unit volume of input space, and therefore is given by the
inverse Jacobian of the mapping
from input space to neural layer:
$M=|J|^{-1}=|\det({{d}}{\bf w}/{{d}}{\bf r})|^{-1}$.
(In the following we
consider the case of noninverting mappings, where
$J$ is positive.)
The magnification factor is a property of the networks' response
to a given probability density of stimuli $P({\bf v})$.
To evaluate $M$ in higher dimensions, one in general has to
compute the  equilibrium state of the whole network using 
global knowledge on $P({\bf v})$.

For one-dimensional mappings (and possibly for special geometric
cases in higher dimensions) the magnification factor may follow
an universal magnification law, i.e.{}
$M(\bar{{\bf w}}({\bf r}))$ is a function only of the local
probability density $P$ and independent of both  location
${\bf r}$ in the neural layer and 
$\bar{{\bf w}}({\bf r})$ in input space.

An optimal map from the view of information theory would
reproduce the input probability exactly
($M\sim P({\bf v})^{\rho}$ with $\rho=1$), according
to a power law with exponent~1,
equivalent to
all neurons in the layer fire with same
probability. An algorithm of maximizing mutual information
has been given by Linsker \cite{linsker89}.

For the classical Kohonen algorithm the magnification law
(for one-di\-men\-sio\-nal mappings) is
a power law $M(\bar{{\bf w}}({\bf r}))
\propto P(\bar{{\bf w}}({\bf r}))^{\rho}$ with exponent
$\rho=\frac{2}{3}$ 
\cite{ritter86}.
For a discrete neural layer and especially 
for neighborhood kernels with different shape and 
range there are corrections to the magnification law
\cite{ritter92,dersch,ritter91}.

\clearpage
\fuss{
\noindent \small \markfont  942 \hspace*{0.3cm} J.C. Claussen and H.G. Schuster \hfill\normalsize\rm
}
\section{Magnification Exponent of the Elastic Net}
The necessary condition for the final state
of algorithm~(\ref{eq:elnetseriell}) is that for all neurons
${\bf r}$ the expectation value of the learning step vanishes:
\begin{eqnarray}
\forall_{{\bf r}{\in}R} \;\;\;\; 0 =
\int{\rm d}{\bf v}\;p({\bf v})\delta{\bf w}_{{\bf r}}({\bf v}).
\label{eq:notwend}
\end{eqnarray}
Since this expectation value is equal to the learning step of
the pattern parallel rule~(\ref{eq:elnetseriell}),
equation~(\ref{eq:notwend}) is the
stationary state condition for {\em both} serial and parallel
updating.
Inserting the learning rule~(\ref{eq:elnetseriell}) to
condition~(\ref{eq:notwend}), we obtain for the invariant density
$\bar{w}_r$ in the one-dimensional case:
\begin{eqnarray}
0 = \int
\Big(
(v-\bar{w}_{r})
\frac{{e}^{-(v-\bar{w}_{r})^2/2\sigma^2}
    }{\int {{d}}r^{'}
      {e}^{-(v-\bar{w}_{r^{'}})^2/2\sigma^2}    }
      + \kappa
\mbox{\boldmath$\bigtriangleup$\unboldmath}
      \bar{w}_r
\Big)  P(v) {{d}}v. \nonumber
\end{eqnarray}

In the limit of a continuous neural layer for every stimulus $v$
there exists one unique center of excitation $s$ with $v=w_s.$
Thus we can substitute integration over ${{d}}v$
by integration over ${{d}}s$.
Using the Jacobian $J(s):={{d}}\bar{w}(s)/{{d}}s$, we have
\begin{eqnarray}
0 = \int
&\bigg( &
(\bar{w}(s)-\bar{w}(r))
\frac{{e}^{-(\bar{w}(s)-\bar{w}(r))^2/2\sigma^2}
    }{\int {{d}}r^{'}
   {e}^{-(\bar{w}(s)-\bar{w}(r^{'}))^2/2\sigma^2} }
+ \kappa
\mbox{\boldmath$\bigtriangleup$\unboldmath}
\bar{w}(r) \bigg)
P(\bar{w}(s)) J(s) {{d}}s.\nonumber
\end{eqnarray}
The second term becomes $\kappa \frac{{\rm d}J}{{\rm d}r}$.
The normalization integral is
($p:=s-r^{'}$):
\\ 
\mbox{}~~
$
\int {e}^{-(\bar{w}(s)-\bar{w}(r^{'}))^2/2\sigma^2}
{{d}}r^{'}
=
 \int {e}^{-p^2/2(\sigma/J(s))^2} {{d}}p
+ o(\sigma^3) 
=
 \sqrt{2\pi} \cdot   \frac{\sigma}{J(s)} + o(\sigma^3).
$
\\
For the following equations, 
we define the abbreviation $\bar{P}(r):=P(\bar{w}(r))$.
Using parametric differentiation, substitution 
${\rm{}d}s={\rm{}d}w_s/({\rm{}d}w_s/ds)={\rm{}d}w_s/J(s)$, 
and saddlepoint expansion (method of steepest descent)
for $\sigma\to{}0$, the first integral becomes
(after Simic \cite{simiccomm}):
\begin{eqnarray}
 \frac{1}{\sqrt{2\pi}\;\;\sigma}
&\cdot & 
\int
(\bar{w}(s)-\bar{w}(r))
{e}^{-(\bar{w}(s)-\bar{w}(r))^2/2\sigma^2}
P(\bar{w}(s)) J(s)^2 {{d}}s 
\label{integralparametric}
\nonumber
\\
\nonumber
&=&\frac{\sigma}{\sqrt{2\pi}} \frac{1}{J(r)} \frac{d}{dr}
\int {e}^{-(\bar{w}(s)-\bar{w}(r))^2/2\sigma^2}
P(\bar{w}(s)) J(s)^2 {{d}}s\\
\nonumber
&=&\frac{\sigma}{\sqrt{2\pi}} \frac{1}{J(r)} \frac{d}{dr}
\int {e}^{-(\bar{w}(s)-\bar{w}(r))^2/2\sigma^2}
P(\bar{w}(s)) J(w(s))  {{d}}w(s)\\
&=&\sigma^2 \frac{1}{J(r)} \frac{d}{dr}(\bar{P}(r) J(r) )
 +o(\sigma^4)
=
\sigma^2( \frac{d\bar{P}}{dr}+\frac{\bar{P}}{J}\frac{dJ}{dr})
 +o(\sigma^4). 
\end{eqnarray}
Neglecting higher orders of $\sigma$, we obtain
\begin{eqnarray}
0 = \frac{\sigma^2}{J(r)} \cdot
\frac{{{d}}}{{{d}}r}\Big(\bar{P}J
  + \kappa \frac{{{d}}J}{{{d}}r} \Big).
\label{eq:elnet_pjr}
\end{eqnarray}
This is a first-order nonlinear differential equation for $J(r)$
to a given input density $P(\bar{r})$.
However, this can be expressed explicitly only if (additional to
$P(v)$) the complete equilibrium state $\bar{w}(r)$ is known, and
then one obtains $J(r)$ directly by evaluating the first
derivative. Thus the differential equation~(\ref{eq:elnet_pjr}) gives
further insight only if $J(r)$ follows an universal scaling law
without explicit dependence on the location $r$, that is, $J$
is a function of $\bar{P}$ only.

\clearpage 
\fuss{
\noindent \small \markfont \hfill  Asymptotic Level Density of the Elastic Net \hspace*{0.3cm} 943 \normalsize\rm \\ \\
}
The ansatz $J(r) = J(\bar{P}(r))$ leads for all $r$, where
${{{d}}\bar{P}}/{{{d}}r}\neq 0$, to the
differential equation for the invariant state of the
one-dimensional Elastic Net Algorithm
\begin{eqnarray}
\frac{{{d}}J}{{{d}}\bar{P}} =
- \frac{J}{\bar{P}}
\cdot  \Big(1 +  \frac{\kappa}{\sigma^2}
\frac{J}{\bar{P}}\Big)^{-1}.
\label{eq:elnetdgl}
\end{eqnarray}
The first derivative depends only on $J/\bar{P}$.
%
%
%
%
The gradient field of
(\ref{eq:elnetdgl}) has two regimes:
 For $\kappa/\sigma^2\rightarrow 0$
                 (`soft string tension')
     ${{d}}J/{{d}}\bar{P}
     = -J/\bar{P},$
     therefore $M=J^{-1}\sim P(v)^{1}.$
     The magnification exponent is asymptotically 1
     and cortical representation is near to  
     the optimum given by information theory.
  For $\kappa/\sigma^2\rightarrow \infty$
                    (`hard string tension')
     ${{d}}J/{{d}}\bar{P} \rightarrow 0,$
     therefore $M=J^{-1}$ has a constant value.
     Here all adaptation to the stimuli vanishes, equivalent to a
     magnification exponent of zero.

\begin{figure}[H]
\epsfig{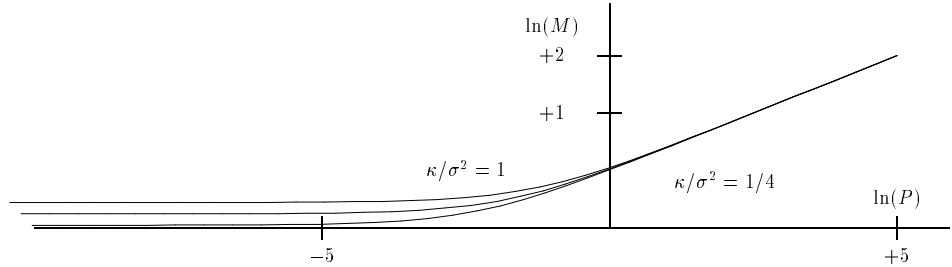}
\vspace*{-3ex}
\caption{Solutions of equation~(\protect\ref{eq:elnetdgl})
for 
 $\kappa/\sigma^2$ =  1, 1/2 (middle) and 1/4.
\label{fig:elnetloesungen}}
\vspace*{-3ex}
\end{figure}

Substituting
$X:=\ln P$, $Y:=-\ln J$ and $Z:=X+Y,$
(\ref{eq:elnetdgl}) can be solved exactly
(see Fig.~\ref{fig:elnetloesungen})
\begin{eqnarray}
\ln M  = \frac{1}{2} \left( \ln(PM) + \ln \big(1 + \frac{1}{2}
  \frac{\kappa}{\sigma^2} \frac{1}{PM}\big)\right) + {\rm const}.
\end{eqnarray}
\vspace*{-3ex}
%
%
%
Thus the magnification 
exponent depends only on the local input probability density
$M\sim P^{\rho(P)}$, and we have
$\rho_q=\frac{{{d}}Y}{{{d}}X}=\rho
+ X\frac{{{d}}\rho(X)}{{{d}}X}$,
where $\rho=\rho_q$ for limiting cases with
${{d}}\rho(X)/{{d}}X\rightarrow 0$.
For $\kappa\to{}0$ the magnification exponent shifts from $1$
to zero according to equation~(\ref{eq:elnetdgl}), 
rewritten as
\begin{eqnarray}
\frac{1}{\rho_q} = \frac{{{d}}X}{{{d}}Y}
= (1 +  \frac{\kappa}{\sigma^2} {e}^{-Z})
= (1 +  \frac{\kappa}{\sigma^2} \frac{1}{\bar{P}M}).
\label{eq:elnetdglloglog}
\end{eqnarray}

Finally we remark that the decomposition 
(\ref{eq:elnetseriell}) of the parallel
update rule to update responses to the stimuli
is not unique.
Especially the elastic term can be decomposed in a siutable
stimulus-dependent manner so that elasticity is appended only in 
vicinity of the stimulus. 
This Local Elastic Net reads
\begin{eqnarray}
\delta {\bf w}_{{\bf r}} &=& \eta  \cdot \{
A^{\sigma}({\bf v},{\bf w}_{{\bf r}}) \cdot
({\bf v} - {\bf w}_{{\bf r}}) 
+ \kappa  ( (1-\nu) \cdot
A^{(\alpha\sigma)}({\bf v},{\bf w}_{{\bf r}}) + \nu ) \cdot
\mbox{\boldmath$\bigtriangleup$\unboldmath}
{\bf w}_{{\bf r}} \}, 
\nonumber
\label{eq:localelnet}
\end{eqnarray}
where $A$ is a normalized gaussian function of distance, 
$\alpha\simeq{}1$
and $0\leq{}\nu\leq{}1$.
A small global elasticity (e.g. $\nu=0.05$) smoothes fluctuations,
but the ``forgetting'' due to global relaxation is reduced
which improves convergence.
The Magnification law of the Local Elastic Net is similar
as for the Elastic Net \cite{gruel92}.

\clearpage
\fuss{
\noindent \small \markfont 944 \hspace*{0.3cm} J.C. Claussen and H.G. Schuster \hfill\normalsize\rm
}
\section{Numerical Verification of the Magnification Law}
\vspace*{-2mm}
To calculate the asymptotic level density numerically,
we considered the
map of the unit interval to a 
onedimensional neural chain of 100 neurons
with fixed first and last neuron.
The learning rate was $0.5$.
The stimulus probability density
was chosen exponentially as $\exp(-\beta{}w)$ with $\beta=4$.
After an adaptation process of $5\cdot{}10^7$ steps
 further $10\%$ of learning steps
were used to calculate average slope and its fluctuation
(shown in brackets)
of $\log{}J$ as a function of $\log{}P.$ (The first and last $10\%$
of neurons were excluded to eliminate boundary effects).
The (local) magnification exponents
were obtained as
\\ \mbox{} \hfill
\newcommand{\myklein}{\footnotesize}
\begin{tabular}[t]{|l|l|l|l|l|}
\hline
$\downarrow\sigma$~~~$\kappa\rightarrow$
& ~~~~ 0.24 ~~~~&~~~ 0.024 ~~~&~~  0.0024 ~~&~ 0.00024 ~\\ \hline
0.0003 ~~~~& 0.00 (0.01)& 0.03 (0.01)& 0.15 (0.02)& 0.29 (0.03)\\
\hline 0.001 ~~~~& 0.03 (0.01)& 0.03 (0.01)& 0.15 (0.02)& 0.28 (0.02)\\
\hline  0.003 ~~~~& 0.04 (0.01)& 0.03 (0.01)& 0.23 (0.01)& 0.49 (0.01)\\
\hline  0.01 ~~~~& 0.03 (0.01)& 0.25 (0.01)& 0.77 (0.02)& 0.96 (0.06)\\
\hline  0.03 ~~~~& 0.23 (0.01)& 0.70 (0.03)&  &  
\normalsize\\\hline
\end{tabular}
\hfill \mbox{}  \\
For the Elastic Net the parameter choice appeared 
crucial:
Same as in the TSP application \cite{simmen91} 
the optimal choice of $\sigma$ as 
the average distance (in input space) between two adjacent neurons 
seems to be appropriate. For larger $\sigma$ clearly clustering
phenomena appear due to the fact that too many neurons 
fall in the Gaussian neighborhood of the stimulus.
For large $\kappa/\sigma^2$ the exponent decreases to zero,
as given by the theory. 
For small $\kappa/\sigma^2$ the exponent first increases near to
$1$ but
simultaneously instability due to clustering arises 
(last row).

Whereas the simulation validates the exact result, appropriate adjustment 
of $\kappa/\sigma^2$ between optimal mapping and stability remains difficult
and becomes intractable for large-scale variations of the
input probability density.

\end{document}